\documentclass[amsmath,floatfix,twocolumn,superscriptaddress,citeautoscript]{revtex4-1}
\usepackage{subfigure}
\usepackage{siunitx}
\usepackage{amssymb}
\usepackage{amsmath}
\usepackage{graphicx}
\usepackage{array}
\usepackage{dcolumn}
\usepackage{psfrag}
\usepackage{bm}
\usepackage{color}
\usepackage{multirow}

\setcitestyle{super}
\newcommand{\inlinecite}[1]{%
  \begingroup
    \bibpunct{}{}{,}{n}{}{}%
    [\citenum{#1}]%
  \endgroup}

\begin{document}

\title{Hydrogen-Bond Density Drives the Order–Disorder Phase Transition from Orientational to Vibrational in Polyalcohol 
       Plastic Crystals}

\author{Ares Sanuy}
    \affiliation{Departament de Física, Universitat Politècnica de Catalunya, 08034 Barcelona, Spain}
    \affiliation{Research Center in Multiscale Science and Engineering, Universitat Politècnica de Catalunya,
                 Campus Diagonal-Besòs, Av. Eduard Maristany 10–14, 08019 Barcelona, Spain}

\author{Carlos Escorihuela-Sayalero}
    \affiliation{Departament de Física, Universitat Politècnica de Catalunya, 08034 Barcelona, Spain}
    \affiliation{Research Center in Multiscale Science and Engineering, Universitat Politècnica de Catalunya,
                 Campus Diagonal-Besòs, Av. Eduard Maristany 10–14, 08019 Barcelona, Spain}

\author{Pol Lloveras}
    \affiliation{Departament de Física, Universitat Politècnica de Catalunya, 08034 Barcelona, Spain}
    \affiliation{Research Center in Multiscale Science and Engineering, Universitat Politècnica de Catalunya,
                 Campus Diagonal-Besòs, Av. Eduard Maristany 10–14, 08019 Barcelona, Spain}

\author{María del Barrio}
    \affiliation{Departament de Física, Universitat Politècnica de Catalunya, 08034 Barcelona, Spain}
    \affiliation{Research Center in Multiscale Science and Engineering, Universitat Politècnica de Catalunya,
                 Campus Diagonal-Besòs, Av. Eduard Maristany 10–14, 08019 Barcelona, Spain}

\author{Hassan Allouchi}
    \affiliation{Synthèse et Isolement de Molécules BioActives, Laboratoire de Chimie Physique, University of Tours, 
                 Av. Monge 31, 37200 Tours, France}

\author{Josep-Lluís Tamarit}
    \affiliation{Departament de Física, Universitat Politècnica de Catalunya, 08034 Barcelona, Spain}
    \affiliation{Research Center in Multiscale Science and Engineering, Universitat Politècnica de Catalunya,
                 Campus Diagonal-Besòs, Av. Eduard Maristany 10–14, 08019 Barcelona, Spain}

\author{Luis Carlos Pardo}
\email{luis.carlos.pardo@upc.edu}
    \affiliation{Departament de Física, Universitat Politècnica de Catalunya, 08034 Barcelona, Spain}
    \affiliation{Research Center in Multiscale Science and Engineering, Universitat Politècnica de Catalunya,
                 Campus Diagonal-Besòs, Av. Eduard Maristany 10–14, 08019 Barcelona, Spain} 

\author{Claudio Cazorla}
\email{claudio.cazorla@upc.edu}
    \affiliation{Departament de Física, Universitat Politècnica de Catalunya, 08034 Barcelona, Spain}
    \affiliation{Research Center in Multiscale Science and Engineering, Universitat Politècnica de Catalunya,
                 Campus Diagonal-Besòs, Av. Eduard Maristany 10–14, 08019 Barcelona, Spain}
    \affiliation{Institució Catalana de Recerca i Estudis Avançats (ICREA), Passeig Lluís Companys 23, 08010 Barcelona, Spain}

\begin{abstract}
\textbf{Abstract.}~Neopentane-derived plastic crystals undergo order-disorder phase transitions accompanied by exceptionally large 
entropy changes, making them attractive solid-state materials for thermal energy storage and barocaloric cooling. Yet it remains 
unclear whether the microscopic nature of this entropy change is an intrinsic property of each compound or can be understood generally 
and tuned systematically through molecular design. Here, we use classical molecular dynamics simulations of neopentyl glycol (NPG), 
pentaglycerine (PG), and pentaerythritol (PE), validated against experimental transition temperatures, enthalpy changes, and phase 
diagrams, to show that the character of the phase-transition entropy undergoes a complete inversion as the hydrogen-bond connectivity 
of the molecule increases: orientational disordering dominates in NPG, whereas lattice vibrations dominate in PE and become remarkably 
robust against pressure. PG occupies a balanced intermediate regime in which the orientational, conformational, and vibrational 
channels contribute comparably. This inversion reflects two coupled microscopic mechanisms: a growing low-frequency gap between 
the phonon densities of states of the ordered and disordered phases as the hydrogen-bond network densifies, and an increase in 
molecular symmetry that progressively suppresses orientational disordering. Because the vibrational channel is substantially less 
sensitive to pressure than the orientational one, hydrogen-bond density emerges as a practical, chemically accessible design handle 
for tailoring not only the magnitude but also the pressure robustness of the caloric response in hydrogen-bonded plastic crystals, 
with direct implications for the search for new barocaloric and thermal-energy-storage materials.
\\

\end{abstract}

\maketitle

\section{Introduction}
\label{sec:intro}
Neopentane-derived polyalcohol plastic crystals undergo order-disorder (OD) phase transitions with exceptionally large entropy changes. 
This property has long made them attractive as solid-state phase-change materials for latent-heat thermal energy storage and conversion 
\cite{introstor1,introstor2,introstor3,introstor4}. More recently, these same transitions have been shown to drive colossal barocaloric 
effects, positioning plastic crystals as promising solid-state alternatives to conventional vapour-compression refrigeration, which relies 
on environmentally harming fluids \cite{introbaro1,introbaro2,introbaro3,expt1,li19,cazorla19}. Neopentyl glycol (NPG) was the first 
member of this family found to display such effects, which arise from the sensitivity of its hydrogen-bonded molecular network to pressure 
\cite{expt1,li19,cazorla19}; the finding has since been extended to related plastic crystals \cite{expt2,intro1,intro2,intro3,intro4}. 
Because the hydrogen-bonded molecular network also plays a crucial role in the orientational dynamics of plastic crystals 
\cite{li20,li22,hbintro1,hbintro2}, the fundamental origin of the OD phase transition has been predominantly attributed to these two 
intertwined factors.

\begin{figure*}
    \centering
       \includegraphics[width=1.00\textwidth]{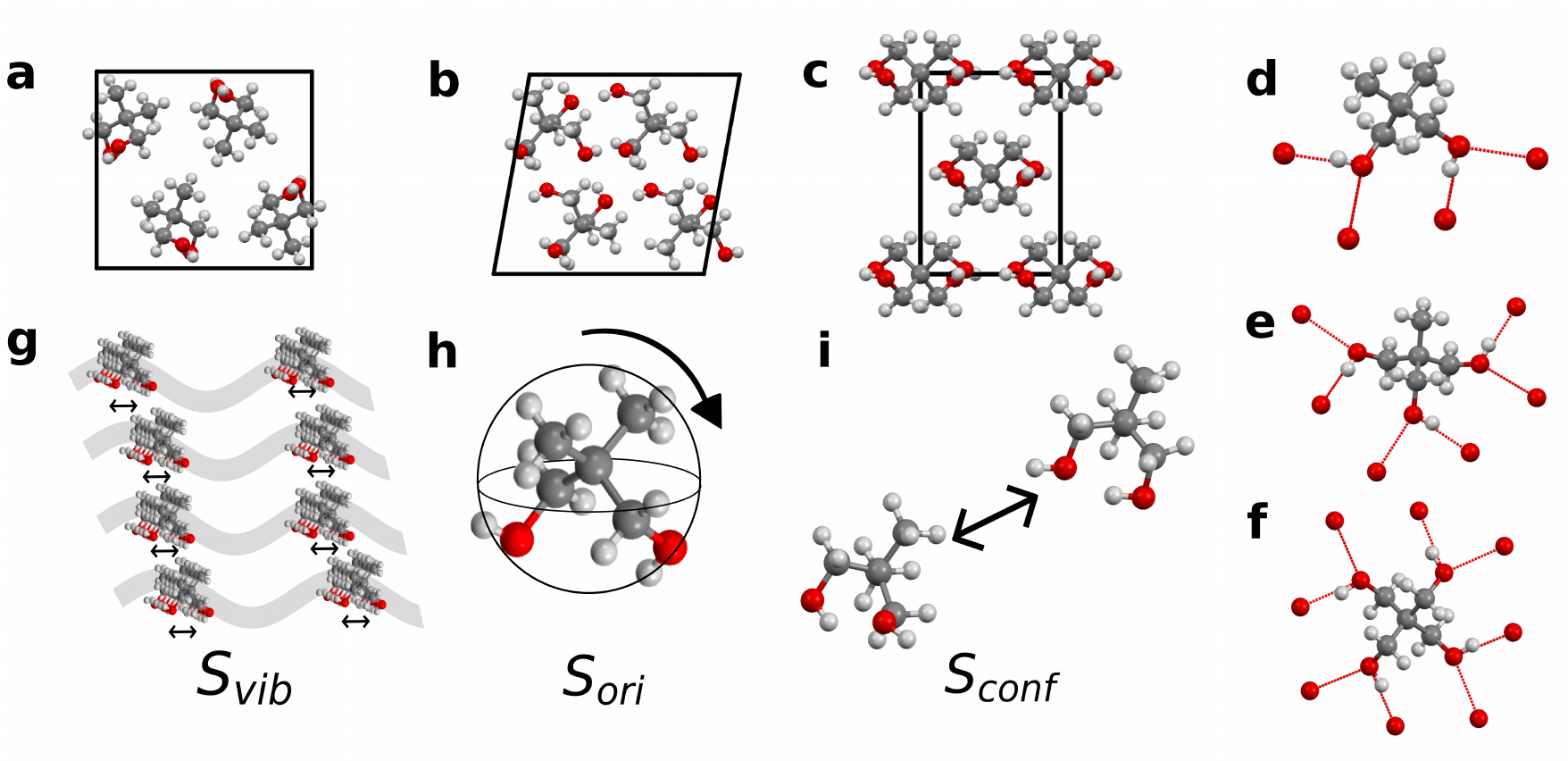}
        \caption{\textbf{Systems description and entropy decomposition.}~Ball-stick representation of the ordered phase of
        (a)~NPG, (b)~PG and (c)~PE; the black solid lines represent the corresponding unit cells. Sketch of the (maximum) number
        of hydrogen bonds (red dotted lines) formed between neighbouring molecules in the ordered phases of (d)~NPG, (e)~PG and
        (f)~PE. Decomposition of the total entropy into (g)~vibrational, (h)~orientational and (i)~conformational contributions.}
        \label{fig1}
\end{figure*}

Despite this progress, the prevailing picture of the OD transition remains largely qualitative: it identifies orientational dynamics
and hydrogen bonding as the key ingredients, but not how they translate into the phase-transition entropy, $\Delta S_{t}$. A central
open question for this class of materials is therefore whether the microscopic nature of $\Delta S_{t}$ is an incidental property of
each individual compound or follows a predictable trend that can be exploited for molecular design. The question extends beyond academic
interest: if the balance among orientational, conformational, and vibrational contributions to $\Delta S_{t}$ can be understood and
systematically tuned, it may offer efficient routes to engineering not only the magnitude of the thermal response but also its robustness
against the external stimulus these materials are designed to exploit, in the present case hydrostatic pressure. A first step toward 
decomposing $\Delta S_{t}$ was recently taken for NPG, where lattice vibrations were shown to play a far greater role than previously 
assumed \cite{sanuy25}. A single compound, however, cannot establish whether this is a generic feature of hydrogen-bonded plastic crystals
or a peculiarity of NPG, nor whether the vibrational contribution grows, shrinks, or remains constant as the hydrogen-bond network
is modified.

In this study, we answer this question directly by studying the NPG--PG~(pentaglycerine)--PE~(pentaerythritol) series, in which the
number of hydroxyl groups per molecule steadily rises from $2$ in NPG to $4$ in PE. This progression provides a natural, chemically
minimal handle on hydrogen-bond connectivity within an otherwise closely related family of molecules. Combining classical molecular
dynamics simulations, information theory, and thermodynamics, we decompose the transition entropy of all three compounds into orientational,
conformational, and vibrational contributions, and we track how this balance evolves with hydrogen-bond density and applied pressure.
We further report an X-ray refinement of the ordered phase of PG, which reveals a triclinic structure (space group $P1$) rather than
the tetragonal symmetry previously assumed.

Our central finding is that the character of the OD transition inverts across the series. Orientational disordering dominates in NPG,
whereas lattice vibrations become the dominant and most pressure-robust contribution in PE, with PG occupying a balanced intermediate
regime in which the three channels contribute nearly equally. A frequency-resolved analysis of the vibrational density of states traces
this inversion to a low-frequency entropy gap between the ordered and disordered phases, which widens systematically as the hydrogen-bond
network densifies. This result establishes a direct microscopic link between hydrogen-bond connectivity and collective lattice dynamics.
Our conclusions recast lattice vibrations from a minor correction to a central, tunable driver of OD transitions in hydrogen-bonded
plastic crystals, and offer a new design principle for barocaloric and thermal-energy-storage materials.

\begin{table*}
\centering
\begin{tabular}{lcccccccc}
\hline
\hline
 & \multicolumn{2}{c}{NPG} &  & \multicolumn{2}{c}{PG} & & \multicolumn{2}{c}{PE} \\
 & This work & Experiment \cite{expt1} &  
 & This work & Experiment \cite{expt2} &  
 & This work & Experiment \cite{expt3} \\ 
\hline
$T_t$ (K)                             & $303 \pm 5$  & $315 \pm 3$  & & $373 \pm 5$ & $354 \pm 3$ & & $477 \pm 5 $ & $459 \pm 3$ \\
$\mathrm{d}T_t/\mathrm{d}P$ (K/GPa)   & $125 \pm 10$ & $113 \pm 5$  & & $96  \pm 7$ & $94  \pm 3$ & & $101 \pm 6 $ &      --     \\
$\Delta H_{t}$ (kJ/kg)                & $130 \pm 5 $ & $121 \pm 2$  & & $170 \pm 5$ & $176 \pm 4$ & & $263 \pm 11$ & $277$       \\
\hline
\hline
\end{tabular}
\caption{\textbf{Data for the OD phase transition in NPG, PG and PE.}~Experimental and theoretical values for the phase-transition 
         temperature, $T_{t}$, OD transition-line pressure slope, $\mathrm{d}T_t/\mathrm{d}P$, and phase-transition enthalpy change, 
         $\Delta H_{t}$, correspond to zero-pressure conditions. Experimental $T_{t}$ uncertainties are derived from the variation 
         among literature values.}
\label{table1}
\end{table*}

\section{Results}
\label{sec:results}
We first introduce the structural properties of the NPG--PG--PE family and present our X-ray refinement of the ordered phase of PG 
(Sec.~\ref{subsec:systems} and Supplementary Information). Subsequently, we briefly describe the atomistic simulation scheme used 
throughout this work (Sec.~\ref{subsec:MD}). Then, the OD phase transition is examined as captured by our classical MD simulations 
(Sec.~\ref{subsec:od}) and compared with available experimental data for the phase-transition temperature, transition-line pressure 
dependence, and phase-transition enthalpy. Next, we quantify how the different entropy contributions evolve under pressure in 
neopentane-derived plastic crystals (Sec.~\ref{subsec:entropy}). Finally, we resolve the vibrational entropy contribution in frequency 
space (Sec.~\ref{subsec:vdos}) and characterize the hydrogen-bond network (Sec.~\ref{subsec:hydrogen}) for each compound to identify 
the global microscopic origin of the OD phase transition across the NPG--PG--PE series.

\subsection{Neopentane-derived polyalcohols}
\label{subsec:systems}
Neopentyl glycol (NPG, C$_{5}$H$_{12}$O$_{2}$, Fig.~\ref{fig1}a), pentaglycerine (PG, C$_{5}$H$_{12}$O$_{3}$, Fig.~\ref{fig1}b), 
and pentaerythritol (PE, C$_{5}$H$_{12}$O$_{4}$, Fig.~\ref{fig1}c) form a family of neopentane-derived polyalcohols sharing a 
common quaternary carbon core, each bearing an increasing number of hydroxyl groups, $n_{\rm OH} = 2$, 3, and 4, respectively. 
The progressive substitution of methyl groups (CH$_{3}$) by hydroxymethyl groups (CH$_{2}$OH) in going from NPG to PE defines 
the key chemical trend within this family: each additional OH group introduces an extra donor site for intermolecular hydrogen
bonding, so that, in their molecular crystalline form, NPG, PG, and PE can form an increasingly larger number of hydrogen bonds 
per molecule (Figs.~\ref{fig1}d--f). Upon heating, all three molecular crystals undergo a first-order solid-solid phase transition 
from an ordered phase to an orientationally disordered (plastic) face-centered-cubic phase, in which the molecules can rotate 
in addition to vibrating and adopting different conformations (Figs.~\ref{fig1}g--i).

\begin{figure}
    \centering
       \includegraphics[width=0.475\textwidth]{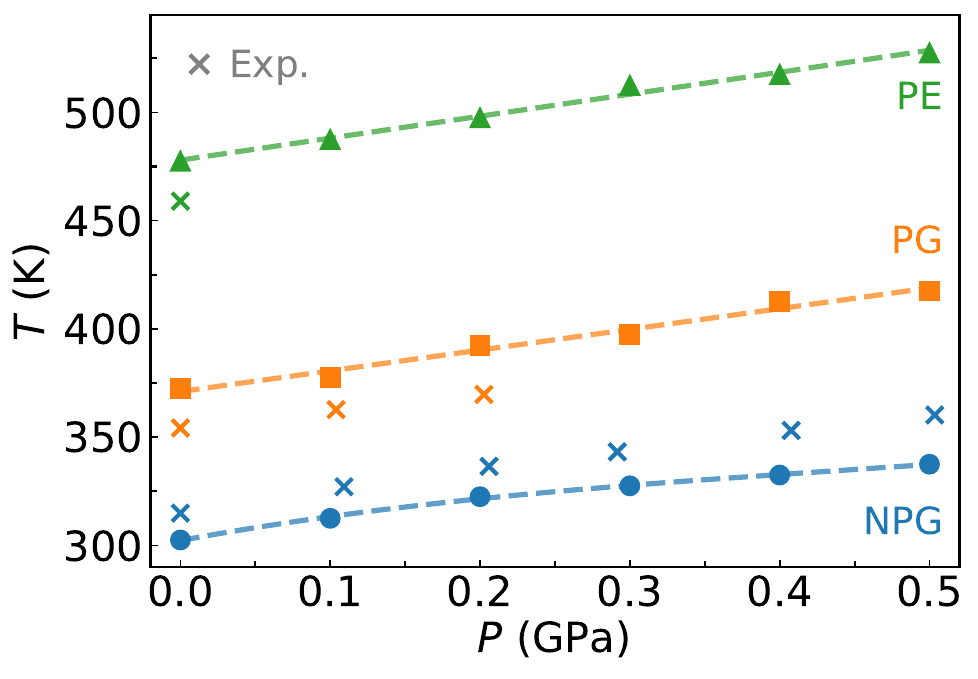}
       \caption{\textbf{Pressure and temperature dependence of the OD phase transition in NPG, PG and PE.}~Simulation results obtained
       in this study are represented with solid symbols and experimental data extracted from works \inlinecite{expt1,expt2,expt3} with
       crosses. Experimental and theoretical uncertainties are smaller than the size of the symbols (Table~1).}
\label{fig2}
\end{figure}

At low temperatures, NPG exhibits an ordered monoclinic $P2_{1}/c$ phase with cell parameters $a = 5.98~(0.02)$, $b = 11.00~(0.03)$ 
and $c = 10.81~(0.03)$~\AA, and angles $\alpha = \beta = 90^{\circ}$ and $\gamma = 112^{\circ}$ \cite{npgexpt}. PE displays a tetragonal 
ordered phase (space group $I\overline{4}$) with lattice parameters $a = b = 6.087~(0.001)$ and $c = 8.757~(0.002)$~\AA~ ($\alpha =
\beta = \gamma = 90^{\circ}$)\cite{peexpt}. As for PG, the corresponding low-temperature ordered phase has been described in the 
literature as tetragonal \cite{pgexpt,pgexpt2}, in analogy to that of PE. However, due to the fact that the molecular symmetry of PG 
is lower than that of PE, we deemed as necessary to experimentally revisit this structural identification. 

We performed X-ray diffraction experiments on a single crystal of PG ($0.150 \times 0.450 \times 0.580$~mm) grown by slow evaporation 
in a methanol solution at $T = 280$~K. Data were collected on an Xcalibur--2 diffractometer (Oxford Diffraction) using graphite 
monochromated Mo K$\alpha$1 radiation ($\lambda = 0.71073$~\AA). Data acquisition was performed at $150$~K using an Oxford Cryosystems 
nitrogen cryostat (Cryostream Plus). The structure was solved by direct methods and successive Fourier difference analysis with the 
\verb!SHELXS! program (\verb!WINGX!, version 2026.1, 1997) \cite{pg1,pg2}, and was subsequently refined by weighted anisotropic full-matrix 
least-squares methods implemented in the \verb!SHELXL! program \cite{pg3}. The positions of non-hydrogen atoms were refined anisotropically 
whereas those of hydrogen atoms were first calculated and then included as isotropic constraints in the analysis. Thermal ellipsoid drawings 
were obtained with the \verb!ORTEP! program \cite{pg1}. Data collection and data reduction were done with the \verb!CRYSALIPRO-CCD! and 
\verb!RED! softwares \cite{pg4}. The rest of calculations were performed with standard procedures embedded in the \verb!WINGX! suite 
of programs \cite{pg1}. 

The structure of PG was found to be triclinic (space group $P1$) with  lattice parameters $a = 6.0822~(0.0002)$, $b = 10.4855~(0.0004)$
and $c = 10.5098~(0.0004)$~\AA, and angles $\alpha = 99.682~(0.003)^{\circ}$, $\beta = 106.289~(0.004)^{\circ}$ and $\gamma = 
90.551~(0.003)^{\circ}$ (additional structural properties can be found in Supplementary Fig.~1). In this ordered structure, each molecule 
is hydrogen bonded to six neighbouring molecules with characteristic donor-acceptor distances ranging from $2.69$ to $2.71$~\AA~ 
(Supplementary Fig.~2). Detailed X-ray crystallographic information files can be obtained free of charge from the Cambridge 
Crystallographic Data Centre, deposit number $2592185$.

\subsection{Computational methods}
\label{subsec:MD}
We performed classical MD simulations of bulk NPG, PG, and PE under varying pressure ($P$) and temperature ($T$) conditions 
at fixed particle number ($N$). These $NPT$-MD simulations were carried out using the \verb!GROMACS! software \cite{gromacs} 
together with CHARMM-type force fields \cite{charmm}. The equations of motion were integrated using a leap-frog algorithm with 
a time step of $2$~fs. Temperature was controlled with a velocity-rescaling thermostat using a coupling time constant of $0.5$~ps, 
while pressure was maintained around the target value using a Berendsen barostat with a coupling time constant of $5$~ps and 
an isotropic compressibility of $4.5 \cdot 10^{-5}$~bar$^{-1}$. In all the simulations, periodic boundary conditions were applied 
along the three Cartesian directions and large supercells containing a total of $512$ molecules were employed. Long-range 
electrostatic interactions were computed with the particle-mesh Ewald method and a real-space cutoff of $1$~nm, and dispersive 
forces were truncated at $1$~nm.

To determine the OD phase-transition temperature at fixed pressure, $T_{t}$, we performed $NPT$-MD simulations across a wide 
thermal range at intervals of $5$~K. The simulations were initiated at temperatures well below the corresponding $T_{t}$, 
considering the experimentally determined ordered structures. At each state point, the system was first equilibrated for $1$~ns 
under the target temperature and pressure conditions. For state points close to the phase transition, where convergence is slowed 
by critical effects, equilibration runs were extended to up to $10$~ns to ensure proper thermalisation. Production runs lasted 
$1$~ns, during which atomic configurations were saved every $1$~ps, yielding a total of $1,000$ uncorrelated configurations per 
state point. These configurations were subsequently used to compute angular probability densities, histograms, and velocity 
autocorrelation functions for our entropy analysis. Further technical details on our $NPT$-MD simulations and entropy calculation 
approach can be found in previous works \inlinecite{sanuy25,cazorla24,escorihuela24,zeng24}.

\subsection{Order-disorder phase transition in neopentane-derived polyalcohols}
\label{subsec:od}
The order-disorder (OD) phase transition was located, for each compound and pressure, as the temperature at which a discontinuous 
jump in the specific volume occurs upon heating, a clear signature of a first-order phase transition (Supplementary Fig.~3). 
The resulting $T_t (P)$ phase boundaries for NPG, PG, and PE, obtained from our $NPT$-MD simulations, are shown in 
Fig.~\ref{fig2}, together with the corresponding zero-pressure values summarised in Table~1. At $P = 0$, we obtain 
$T_{t} = 303 \pm 5$~K for NPG, $T_{t} = 373 \pm 5$~K for PG, and $T_{t} = 477 \pm 5$~K for PE. The phase-transition 
temperature thus increases monotonically along the series by more than $170$~K from NPG to PE, tracking the progressive 
densification of the hydrogen-bond network as the number of hydroxyl groups per molecule increases from $n_{\rm OH} = 2$ 
to $4$. This trend is consistent with a picture in which a more extensively hydrogen-bonded ordered phase requires a 
correspondingly larger thermal energy to disorder \cite{li20,li22}. 

The pressure dependence of $T_{t}$ also reveals a qualitatively distinct behaviour across the family. For NPG, the $T_{t} (P)$ 
curve is markedly nonlinear and is well described by a third-order polynomial, consistent with previous reports for this 
compound,\cite{expt1} the corresponding slope at $P = 0$ being $dT_{t}/dP = 125 \pm 10$~K/GPa. In contrast, the phase boundaries 
of PG and PE are, within our simulation uncertainty, well described by linear fits over the whole pressure range explored 
($0$--$0.5$~GPa), with slopes $dT_{t}/dP = 96 \pm 7$~K/GPa and $101 \pm 6$~K/GPa, respectively (Table~1). The overall steepness 
of $T_{t} (P)$ is comparable across the three compounds, indicating that, despite their markedly different hydrogen-bond connectivity, 
the sensitivity of the OD transition to applied pressure remains of a similar order of magnitude throughout the family; it is 
instead the curvature of the phase-transition line, most pronounced in NPG, that distinguishes the pressure response of the least 
hydrogen-bonded compound from its more densely bonded counterparts.

Comparison with available experimental data provides a stringent test of our simulation approach. At zero pressure, the simulated 
$T_{t}$ values follow closely the experimental ones for all three compounds (Table~1 and Fig.~\ref{fig2}). The simulated pressure 
slopes show similarly good agreement: $dT_t/dP = 125 \pm 10$~K/GPa versus the experimental $113 \pm 5$~K/GPa for NPG, and $96 \pm 
7$~K/GPa versus $94 \pm 3$~K/GPa for PG. No experimental pressure dependence has been reported for PE to date, so our simulated 
slope of $101 \pm 6$~K/GPa is presented here as a prediction. The phase-transition enthalpies, $\Delta H_{t}$, obtained from our 
$NPT$-MD simulations are likewise in close agreement with observations across the series, falling within or close to the reported 
experimental values for all three compounds (Table~1). Since $\Delta S_{t} = \Delta H_{t} / T_{t}$, this agreement in both $T_{t}$ 
and $\Delta H_{t}$ implies comparably good agreement in the experimental phase-transition entropy.

Taken together, the consistent agreement between simulation and experiment in $T_{t}$, $dT_{t}/dP$, and $\Delta H_t$ across three 
chemically distinct compounds, spanning a $\sim 170$~K range in transition temperature and hydrogen-bond networks of markedly
different density, confirms that our $NPT$-MD approach reproduces the thermodynamics of the OD phase transition with excellent 
reliability across the NPG--PG--PE family. On this basis, we are confident that the outputs of our computational framework 
can be used to reliably decompose the phase-transition entropy into its orientational, conformational, and vibrational contributions, 
as presented in the next section.

\begin{figure*}
    \centering
       \includegraphics[width=1.00\textwidth]{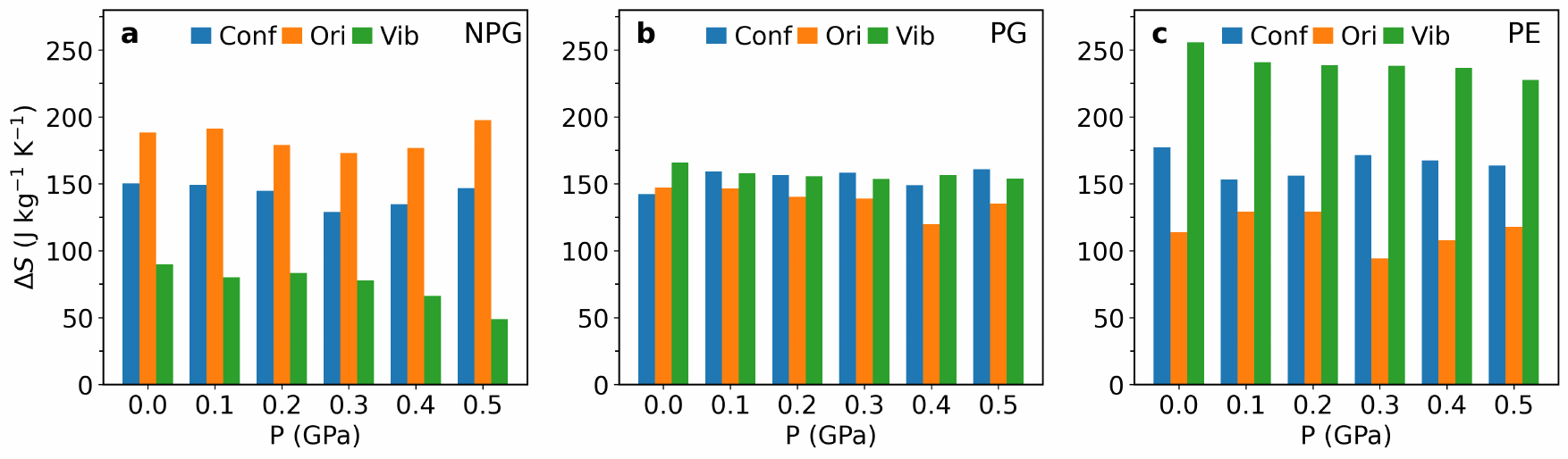}
       \caption{\textbf{Phase-transition entropy change decomposition in neopentane-derived polyalcohol plastic crystals under 
                compression.}~Conformational (blue), orientational (orange), and vibrational (green) contributions are shown 
                for (a)~NPG, (b)~PG, and (c)~PE as a function of pressure.}
        \label{fig3}
\end{figure*}

\subsection{Phase-transition entropy change decomposition}
\label{subsec:entropy}
The OD transition entropy change of a non-metallic, non-magnetic plastic crystal can be decomposed as
\begin{equation}
        \Delta S_{t} = \Delta S_{t}^{\rm vib} + \Delta S_{t}^{\rm ori} + \Delta S_{t}^{\rm conf}~,
\label{eq1}
\end{equation}
where $\Delta S_{t}^{\rm x} \equiv S_{\rm x}({\rm disordered}) - S_{\rm x}({\rm ordered})$, and all entropy terms are 
evaluated at the transition temperature corresponding to a given pressure. Here, $\Delta S_{t}^{\rm vib}$ accounts for 
the change in vibrational (phonon) modes across the phase transition, $\Delta S_{t}^{\rm ori}$ captures the variation 
in the orientational dynamics of the molecules, and $\Delta S_{t}^{\rm conf}$ reflects the change in the range of
molecular conformations adopted across the transition (Figs.~\ref{fig1}g--i). Each of these entropy terms was computed 
following the methods described in our previous works \inlinecite{sanuy25,cazorla24,escorihuela24,zeng24}, to which the 
interested reader is referred for further technical details. For completitude purposes, we enclose the orientational 
and conformational molecular histograms computed for the ordered and disordered phases of NPG, PG and PE in Supplementary 
Figs.~4--14.

Figure~\ref{fig3} shows the pressure dependence of the three phase-transition entropy contributions calculated across the 
NPG--PG--PE family. In NPG, the orientational term dominates at all explored pressures ($\Delta S_{t}^{\rm ori} \approx 
188$--$196$~J\,kg$^{-1}$K$^{-1}$), followed by the conformational contribution ($\Delta S_{t}^{\rm conf} \approx 
129$--$150$~J\,kg$^{-1}$K$^{-1}$), while the vibrational term is the smallest ($\Delta S_{t}^{\rm vib} \approx 
49$--$90$~J\,kg$^{-1}$K$^{-1}$). In PG, the picture changes markedly: the three contributions become remarkably comparable 
in magnitude throughout the analysed pressure range, with $\Delta S_{t}^{\rm conf} \approx 142$--$161$~J\,kg$^{-1}$K$^{-1}$, 
$\Delta S_{t}^{\rm ori} \approx 120$--$147$~J\,kg$^{-1}$K$^{-1}$, and, in stark contrast to NPG, an essentially pressure-independent 
vibrational term, $\Delta S_{t}^{\rm vib} \approx 154$--$166$~J\,kg$^{-1}$K$^{-1}$. This trend continues in PE, where the 
vibrational contribution becomes the clearly dominant term at all pressures, reaching $256$~J\,kg$^{-1}$K$^{-1}$ at zero 
pressure and remaining remarkably stable up to $0.5$~GPa, while the orientational term is now the smallest of the three 
($\Delta S_{t}^{\rm ori} \approx 94$--$130$~J\,kg$^{-1}$K$^{-1}$).

The most striking trend to emerge from Fig.~\ref{fig3} is therefore the pronounced growth of the vibrational contribution 
as the family is traversed from NPG to PE. At ambient pressure, $\Delta S_{t}^{\rm vib}$ increases from $90$~J\,kg$^{-1}$K$^{-1}$ 
in NPG, to $\sim 154$~J\,kg$^{-1}$K$^{-1}$ in PG, and up to $256$~J\,kg$^{-1}$K$^{-1}$ in PE, nearly a threefold increase 
across the series. This is accompanied by a moderate but systematic decrease in the orientational contribution, from 
$\Delta S_{t}^{\rm ori} \approx 190$~J\,kg$^{-1}$K$^{-1}$ in NPG down to $\sim 130$~J\,kg$^{-1}$K$^{-1}$ in PG, and 
$\sim 110$~J\,kg$^{-1}$K$^{-1}$ in PE. The conformational contribution, by contrast, remains comparatively stable across the
entire family, fluctuating only modestly around $130$--$175$~J\,kg$^{-1}$K$^{-1}$ regardless of the compound and pressure.

The decrease in the orientational entropy contribution as one moves from NPG to PG (and, further, to PE) can be rationalised 
in terms of symmetry arguments. As additional CH$_{3}$ groups are substituted by CH$_{2}$OH groups along the series, the symmetry 
of the constituent molecules increases. Since the symmetry of the high-temperature phase is the same in all compounds, the number 
of distinguishable molecular orientations compatible with a face-centered cubic lattice decreases under increasing molecular symmetry 
(Supplementary Figs.~4--14). This reduction in the orientational phase space accessible to the molecules at high temperatures 
implies a decrease in the entropy gained through orientational disordering at the transition. This rationale provides a direct 
mechanical link between the chemical substitution pattern that defines the NPG--PG--PE family and the reduction in 
$\Delta S_{t}^{\rm ori}$ observed in Fig.~\ref{fig3}. The physical origin of the concurrent and much larger increase in the 
vibrational contribution is analysed in detail in Sec.~\ref{subsec:vdos}. 

Beyond the absolute magnitudes, it is instructive to consider the relative weight of each contribution to $\Delta S_{t}$ and how 
this balance shifts with both pressure and compound. At ambient pressure, the orientational term accounts for nearly half of 
$\Delta S_{t}$ in NPG ($44\%$), with the conformational ($35\%$) and vibrational ($21\%$) terms playing secondary roles. In PG, 
the three contributions become nearly equitable, at $36\%$, $32\%$, and $32\%$ for the vibrational, orientational, and conformational 
channels, respectively. In PE, this hierarchy is completely inverted: the vibrational term now dominates ($47\%$), the conformational 
contribution remains close to its value in the other two compounds ($32\%$), and the orientational contribution is reduced 
to only $21\%$ of the total. The conformational contribution thus emerges as the most stable term, consistently accounting for 
$\sim 30$--$37\%$ of $\Delta S_{t}$ across the whole series and pressure range.

This inversion of the entropy hierarchy is also robust against pressure, although the degree of pressure sensitivity itself varies 
markedly between compounds. In NPG, the orientational dominance persists across the full pressure range, but the already modest 
vibrational weight is further depleted under compression, falling from $21\%$ at $P = 0$ to only $13\%$ at $P = 0.5$~GPa. In PG, the 
three relative weights remain within five percentage points of one another at all pressures, confirming that the balanced distribution 
of entropy channels is an intrinsic feature of this compound rather than an artefact of a particular pressure. In PE, by contrast, 
$\Delta S_{t}^{\rm vib}$ accounts for $45$--$47\%$ of $\Delta S_{t}$ across the full pressure range studied, varying by only $11\%$ 
in absolute terms despite the applied compression. Taken together, these results show that, as $n_{\rm OH}$ increases, the vibrational 
channel evolves from being the smallest and most pressure-sensitive contribution in NPG to the dominant and most pressure-stable one 
in PE, a trend whose microscopic origin we investigate next.

\begin{figure*}
    \centering
       \includegraphics[width=1.00\textwidth]{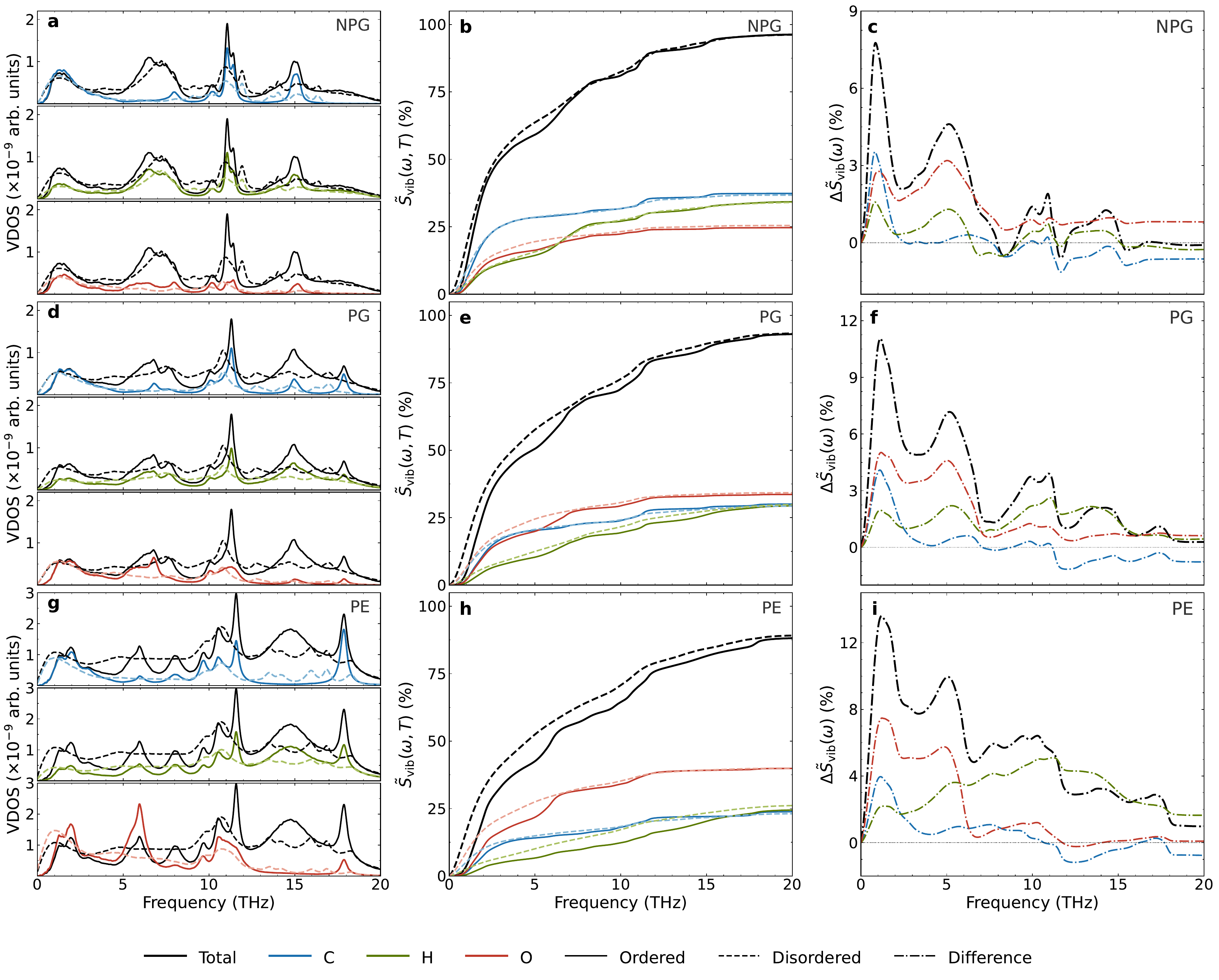}
       \caption{\textbf{Vibrational features of the OD phase transition in neopentane-derived polyalcohol plastic crystals at
                zero pressure.}~Projected density of vibrational states (VDOS), normalized cumulative vibrational entropy, 
                $\tilde{S}_{\rm vib}$, and normalized cumulative vibrational entropy difference between the ordered and 
                disordered phases, $\Delta \tilde{S}_{\rm vib}$, for (a)--(c)~NPG, (d)--(f)~PG and (g)--(i)~PE. Solid and dashed 
                lines in the VDOS and $\tilde{S}_{\rm vib}$ plots correspond to ordered and disordered phases, respectively. 
                All quantities are evaluated at the corresponding phase-transition temperature $T_{t}$.}
        \label{fig4}
\end{figure*}

\begin{figure*}
    \centering
       \includegraphics[width=1.0\textwidth]{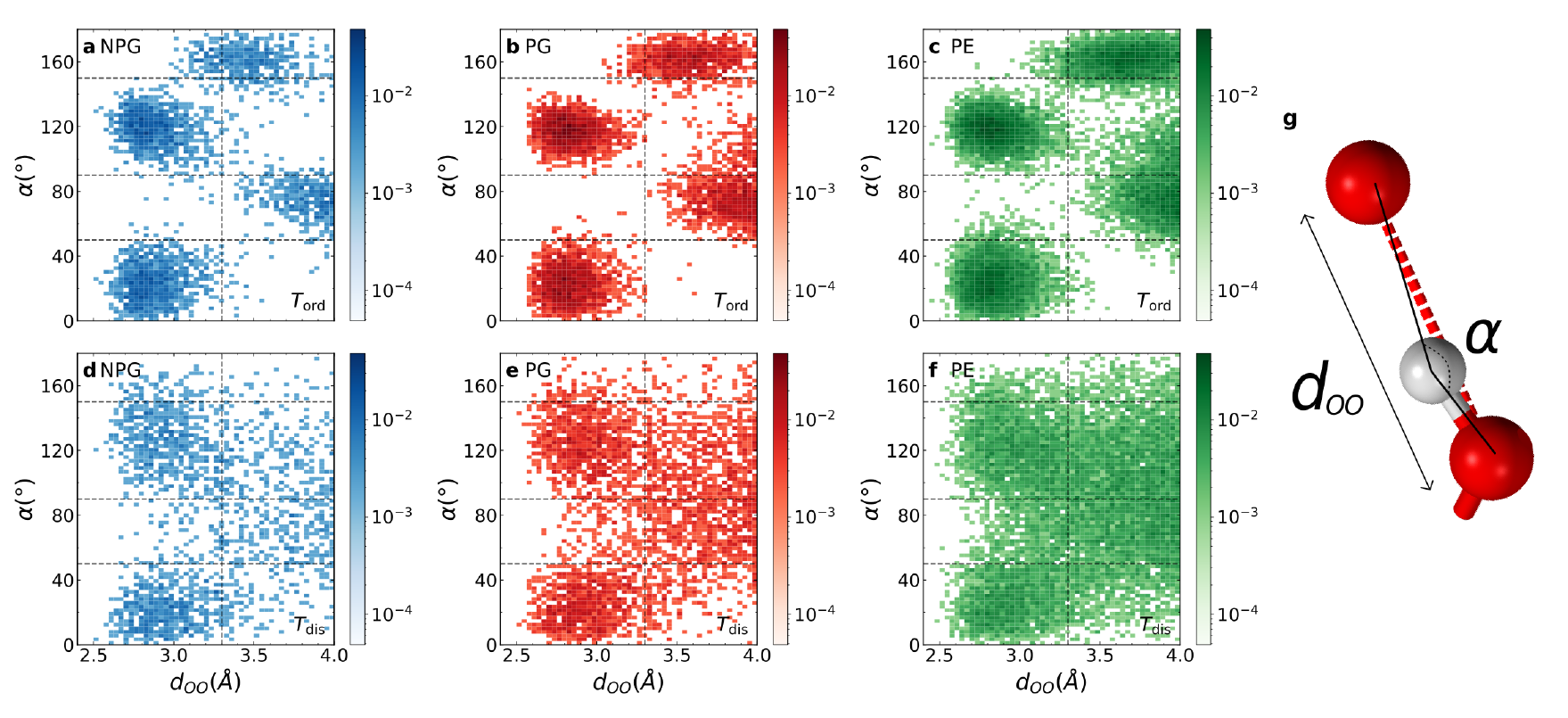}
       \caption{\textbf{Hydrogen-bond network analysis for neopentane-derived polyalcohol plastic crystals at zero 
        pressure.}~Two-dimensional O--H··O angle, $\alpha$, versus oxygen donor-acceptor distance, $d_{\rm OO}$, histograms for the 
        ordered (top row) and disordered (bottom row) phases of (a),(d)~NPG, (b),(e)~PG and (c),(f)~PE. Black discontinuous lines 
        represent the limits employed for estimating the number of hydrogen bonds per molecule, $n_{\rm HB}$. (g)~Schematic 
        representation of the $\alpha$ angle and $d_{\rm OO}$ distance; oxygen and hydrogen atoms are shown as red and white 
        spheres, respectively.} 
        \label{fig5}
\end{figure*}

\subsection{Vibrational density of states analysis}
\label{subsec:vdos}
The systematic inversion of the entropy hierarchy identified above, from an orientation-dominated transition in NPG to a 
vibration-dominated one in PE, raises the question of which specific phonon modes are responsible for this behaviour, and 
how the hydrogen-bond network shapes them. Figure~\ref{fig4} presents a comparative analysis of the vibrational density of 
states (VDOS) and the associated vibrational entropy for the three systems studied, each in their ordered and disordered phases.

Figures~\ref{fig4}a,d,g show the elemental VDOS decomposed by atomic species, together with the total VDOS for each system. 
The ordered and disordered phases exhibit overall similar spectral shapes, but the disordered configurations display systematically 
broader and less sharply resolved peaks across all frequency ranges, reflecting the loss of orientational long-range structural 
periodicity. This broadening is most pronounced in the low-frequency region (below approximately 8~THz), associated with lattice 
and intermolecular collective motions, whereas the high-frequency intramolecular modes (C--H and O--H stretching and bending, above
approximately 12~THz) remain comparatively well defined in both phases, as expected for vibrations governed primarily by local 
covalent bonding rather than long-range order. This contrast in the VDOS becomes more pronounced along the series: the low-frequency 
spectral weight accounts for an increasingly larger fraction of the total VDOS on moving from NPG to PG to PE, consistent with 
the growing number of low-energy lattice modes made available by the denser hydrogen-bond network of the heavier compounds.

Figures~\ref{fig4}b,e,h show the normalised cumulative vibrational entropy as a function of frequency, defined as:
\begin{equation}
        \tilde{S}_{\rm vib}(\omega) \equiv \frac{S_{\rm vib}^{\rm cum}(\omega)}{S_{\rm vib}^{\rm tot}} \times 100~(\%)~,
\label{eq2}
\end{equation}
where all quantities are evaluated at $T_t$, $S_{\rm vib}^{\rm tot}$ represents the total vibrational entropy, and by construction 
$S_{\rm vib}^{\rm cum}(\infty) = S_{\rm vib}^{\rm tot}$. In all three systems, the disordered phase accumulates vibrational entropy 
considerably faster at low frequencies than its ordered counterpart, with the two curves converging only at higher frequencies 
\cite{tamarit1,tamarit2}. This behaviour is consistent with the VDOS broadening observed in Figs.~\ref{fig4}a,d,g: disorder enhances 
the density of low-frequency modes, which in turn dominate the vibrational entropy owing to the larger thermal population of low-energy 
lattice excitations at low and moderate temperatures \cite{sanuy25,cazorla24,escorihuela24,zeng24}. Quantitatively, the disordered 
phase already accounts for roughly half of its total vibrational entropy by $\sim 3$--4~THz in all three compounds, whereas the 
ordered phase reaches the same fraction only at appreciably higher frequencies, an offset that becomes progressively larger from 
NPG to PE and that mirrors the widening of the low-frequency entropy gap discussed below. The two curves ultimately converge above 
$\sim 15$~THz, confirming that the excess entropy of the disordered phase originates almost entirely from the low- and mid-frequency 
spectral range.

Figures~\ref{fig4}c,f,i isolate this effect directly by showing the normalised cumulative vibrational entropy difference:
\begin{equation}
        \Delta \tilde{S}_{\rm vib}(\omega) = \tilde{S}_{\rm vib}^{\rm dis}(\omega) - \tilde{S}_{\rm vib}^{\rm ord}(\omega)~.
\label{eq3}
\end{equation}
This quantity directly measures, at each frequency, the fraction of the total vibrational entropy gained by the disordered phase 
relative to the ordered one. A characteristic double-peak structure emerges in all three systems, with a sharp maximum near 1~THz 
and a secondary, broader maximum near 5~THz, indicating that the entropic difference between order and disorder is concentrated 
almost entirely within the low-frequency lattice and intermolecular modes. The height of the primary, sharpest peak near 1~THz grows
monotonically along the series, from $\sim 7\%$ in NPG to $\sim 11\%$ in PG and $\sim 13\%$ in PE, indicating that the lowest-frequency 
lattice modes become an increasingly effective channel for entropy gain as the hydrogen-bond network densifies. Beyond approximately
12~THz, the curves approach zero and, in some cases (notably the carbon contribution in NPG and PG), become slightly negative, 
suggesting a marginal entropy advantage of the ordered phase at high frequency (although this effect is small and likely close to 
the noise level of the calculation). Oxygen consistently dominates the secondary peak around 5~THz in NPG and PG, and remains the 
largest single elemental contribution to $\Delta \tilde{S}_{\rm vib}$ in PE as well, pointing to a particular sensitivity 
of oxygen-related vibrational modes, plausibly linked to hydrogen-bond network reorganisation, to the degree of structural order.

Comparing across systems, the overall magnitude of the low-frequency vibrational entropy gap increases systematically from NPG 
($\sim 8\%$) to PG ($\sim 11\%$) to PE ($\sim 14\%$), a nearly twofold growth across the series that closely parallels the 
corresponding increase in $n_{\rm HB}$ reported in Table~2. This correspondence suggests that larger and more strongly hydrogen-bonded 
molecular packings exhibit a systematically stronger entropic response to disorder in their collective, low-frequency degrees of 
freedom. This trend indicates that, as the number of hydroxyl groups, and thus the number of hydrogen bonds (Figs.~\ref{fig1}d--f), 
increases along the NPG--PG--PE series, the separation between the ordered and disordered VDOS becomes progressively more pronounced, 
particularly in the low-frequency region. It can therefore be concluded that hydrogen bonding plays a central role in coupling the 
local molecular structure to the collective lattice dynamics. Specifically, a denser hydrogen-bond network amplifies the structural 
sensitivity of the low-frequency vibrational modes to the degree of order, which is consistent both with the systematic increase in 
$\Delta \tilde{S}_{\rm vib}$ and with the nearly threefold growth in the absolute vibrational entropy contribution, $\Delta S_t^{\rm vib}$, 
reported in Sec.~\ref{subsec:entropy} across the same series of neopentane-derived polyalcohols.

\begin{table}
\centering
\begin{tabular}{ccccc}
\hline
\hline
Compound & \quad $n_{\rm OH}$ \quad & \quad $n_{\rm HB}^{\rm ord}$ \quad & \quad $n_{\rm HB}^{\rm dis}$ \quad & $\Delta n_{\rm HB}$ \\
\hline
NPG  & $2$ & $3.94$ & $2.48$ & 1.46  \\
PG   & $3$ & $5.92$ & $3.69$ & 2.23  \\
PE   & $4$ & $7.81$ & $4.29$ & 3.52  \\
\hline
\hline
\end{tabular}
\caption{\textbf{Hydrogen-bond network data for neopentane-derived polyalcohol plastic crystals at zero pressure.}~Number of
         OH groups per molecule, $n_{\rm OH}$, and of averaged hydrogen bonds formed by a molecule, $n_{\rm HB}$, in the ordered
         and disordered phases of NPG, PG and PE. $\Delta n_{\rm HB}$ is defined as the difference
         $n_{\rm HB}^{\rm ord} - n_{\rm HB}^{\rm dis}$.}
\label{table2}
\end{table}

\subsection{Hydrogen-bond network analysis}
\label{subsec:hydrogen}
To characterise the hydrogen-bond (HB) network across the NPG--PG--PE series, we performed a geometric analysis of all 
intermolecular O--H$\cdots$O contacts using the MDAnalysis library \cite{MDAnalysis}. For each pair of oxygen atoms 
belonging to different molecules, we computed the oxygen--oxygen distance, $d_{\rm OO}$, and the O--H$\cdots$O angle, 
$\alpha$ (Fig.~\ref{fig5}g), retaining all contacts with $d_{\rm OO} < 4.0$~\AA, regardless of whether they correspond 
to a genuine hydrogen bond. The analysis was carried out at each state point over several configurations extracted from 
the trajectories generated during the $NPT$--MD simulations. Figure~\ref{fig5} shows the resulting two-dimensional 
distribution of $(d_{\rm OO}, \alpha)$ pairs for the ordered and disordered phases of NPG, PG, and PE, plotting every 
O$\cdots$O pair detected under the criterion above.

In all three compounds, in the ordered state, two well-defined clusters appear at $d_{\rm OO} \approx 2.8$~\AA, one centred 
at $\alpha \approx 120^{\circ}$ and the other at $\alpha \approx 20^{\circ}$. Since each hydroxyl oxygen possesses both 
a covalently bonded hydrogen and an electronic lone pair, it can simultaneously act as donor (through its own H) and as 
acceptor (through the lone pair) within the HB network. As a result, for a given O$\cdots$O pair at the HB distance, the 
angle $\alpha$ can be measured either from the hydrogen of the donor oxygen, giving the $\alpha \approx 120^{\circ}$ 
population, or from the hydrogen of the partner oxygen acting reciprocally as donor, giving the $\alpha \approx 20^{\circ}$ 
population; both therefore correspond to genuine hydrogen bonds, rather than to distinct or spurious populations. The 
additional, more diffuse signal at larger distances ($d_{\rm OO} > 3.3$~\AA) instead arises from non-bonded O$\cdots$O 
contacts between neighbouring molecules that do not constitute genuine hydrogen bonds. 

On this basis, and similarly to previous works \cite{hb1,hb2}, we define a true intermolecular hydrogen bond as any contact 
with $d_{\rm OO} < 3.3$~\AA\ and $\alpha \in [0^{\circ}, 50^{\circ}] \cup [90^{\circ}, 150^{\circ}]$, encompassing both the 
$\alpha \approx 120^{\circ}$ and $\alpha \approx 20^{\circ}$ populations identified in Fig.~\ref{fig5}. The distance threshold 
was quantitatively chosen as the first minimum in the oxygen-oxygen radial distribution function, thereby restricting the analysis 
to first-neighbour O$\cdots$O pairs. The angular threshold, by contrast, was set empirically from the extent of the clusters 
appearing in the $(d_{\rm OO}, \alpha)$ maps, hence is inherently more arbitrary. Although other classification schemes may 
exist \cite{santos25}, the present hydrogen-bond identification criterion is applied identically to the ordered and disordered 
phases of all three compounds, thus ensuring a consistent comparison across the series.  

Using this criterion, the average number of hydrogen bonds per molecule, $n_{\rm HB}$, was computed for both phases and is 
summarised in Table~2. In the ordered phase, $n_{\rm HB}$ increases monotonically along the series, from $3.94$ in NPG to 
$5.92$ in PG and $7.81$ in PE, closely tracking the number of hydroxyl groups per molecule with a ratio of approximately 
two ($n_{\rm HB}^{\rm ord}/n_{\rm OH} \approx 1.97$, $1.97$, and $1.95$ for NPG, PG, and PE, respectively). This near-perfect 
proportionality indicates that essentially every hydroxyl group in the ordered phase participates fully in the HB network, 
acting once as donor and  once as acceptor. Upon the OD phase transition, $n_{\rm HB}$ decreases substantially in all compounds, 
with a relative loss of $37\%$ for NPG, $38\%$ for PG, and $45\%$ for PE; PE also loses the largest absolute number of HBs 
per molecule: $\Delta n_{\rm HB} = 3.52$, versus $1.46$ for NPG and $2.23$ for PG. The fact that PE exhibits both the largest 
relative and the largest absolute loss in $n_{\rm HB}$ indicates that its hydrogen-bond network, despite of being the most 
extensive in the ordered phase, is proportionally more fragile against orientational disordering than those of NPG and PG.

The described HB trend is directly reflected in the shape of the $(d_{\rm OO}, \alpha)$ distribution. In the ordered phase, the 
HB clusters are sharply localised, reflecting the orientational long-range order of the crystal; this localisation is already 
clear for NPG and becomes progressively tighter along the NPG--PG--PE series, in line with the increasing $n_{\rm HB}$ reported 
above. Upon the OD phase transition, the clusters broaden substantially and increasingly overlap with the background of non-bonded 
O$\cdots$O contacts, consistent with the drop in $n_{\rm HB}$. This behaviour can be understood by the fact that, as molecules 
begin to rotate, the angular preference is lost and, consequently, the O$\cdots$O distance distribution broadens significantly. 
(This interpretation is consistent with the elongation of the average HB length reported for the disordered phase of NPG by 
Li~\textit{et al.} \cite{li19}) The cluster overlapping effect is most pronounced for PE, where the two-dimensional distribution 
in the disordered phase is nearly uniform over the entire $(d_{\rm OO}, \alpha)$ space, reflecting the highest degree of thermal 
fluctuations within the family; by contrast, the disordered distributions of NPG and, to a lesser extent, PG still retain visible 
residual density near the ordered-phase cluster positions, indicating a comparatively weaker loss of local directional correlation.  

The progressive loss of HB directionality and connectivity along the NPG--PG--PE series upon the OD phase transition provides a 
direct microscopic picture of why the vibrational entropy contribution grows most strongly in PE: the breakdown of its denser HB 
network releases a larger number of low-frequency modes, which dominate the vibrational density of states in the disordered phase. 
Taken together, the systematic increase in both $n_{\rm HB}$ and $\Delta n_{\rm HB}$ from NPG to PE (Table~2) parallels the 
systematic increase in the vibrational entropy contribution and in the low-frequency vibrational entropy gap reported in 
Secs.~\ref{subsec:entropy} and~\ref{subsec:vdos}. Therefore, hydrogen-bond density emerges as the common microscopic variable 
underlying the inversion of the transition-entropy hierarchy across the NPG--PG--PE family.

\section{Conclusions and Outlook}
\label{sec:conclusions}
We have used classical MD simulations and advanced molecular analysis tools to decompose the OD phase-transition entropy of 
the neopentane-derived polyalcohol family NPG--PG--PE into its orientational, conformational, and vibrational contributions, 
and to follow how this balance evolves as the hydrogen-bond connectivity of the molecule increases. Our $NPT$--MD 
simulations reproduce the experimental phase-transition temperatures, OD phase boundaries, and phase-transition enthalpies 
of all three compounds with notable accuracy, validating the computational framework employed throughout this 
work. Building on this validated approach, we find that the microscopic character of the transition entropy 
is not fixed but undergoes a complete and systematic inversion across the family: orientational disorder 
dominates $\Delta S_t$ in NPG, lattice vibrations dominate in PE, and PG occupies a genuinely balanced intermediate 
regime in which all three channels contribute comparably. This inversion is accompanied by a threefold increase 
in the vibrational contribution and a moderate decrease in the orientational contribution as the series is 
traversed, while the conformational term remains remarkably stable in both absolute and relative terms. 

Analysis of the vibrational density of states and hydrogen-bond network trace the growth of the vibrational 
contribution to a low-frequency entropy gap between the ordered and disordered phonon spectra that widens 
systematically as the number of hydrogen bonds per molecule increases. The breakdown of the hydrogen-bond 
network at the phase transition releases an increasing number of low-frequency lattice and intermolecular 
vibrational modes as $n_{\rm OH}$ grows, directly coupling the local chemistry of the molecule, namely the 
number and connectivity of its hydroxyl groups, to the collective lattice dynamics of the crystal. Complementarily, 
the decline of the orientational contribution across the series is rationalised by the concurrent increase 
in molecular symmetry as methyl groups are progressively replaced by hydroxymethyl groups, which reduces the 
orientational phase space accessible to the molecules at high temperatures and thus their gain in orientational 
entropy. Taken together, these two mechanisms, a stiffening hydrogen-bond network that reshapes the phonon 
spectrum and an increase in molecular symmetry that diminishes orientational disorder, provide a coherent, 
microscopically grounded explanation for the inversion of the entropy hierarchy observed across the NPG--PG--PE 
family.

These results carry direct significance beyond the specific compounds studied here. For the broader community 
working on plastic crystals and hydrogen-bonded molecular solids, our findings establish that lattice vibrations 
are not a minor correction to the OD phase-transition entropy but can become its dominant component once the 
hydrogen-bond network is sufficiently dense. This reframes the vibrational channel as a central driver of entropy 
change in this class of materials, on par with the orientational disordering that has traditionally received the 
most attention. For the energy-storage and barocaloric-cooling communities in particular, the finding that the 
vibrational contribution is markedly more robust to applied pressure than the orientational one identifies 
hydrogen-bond density as a practical, chemically accessible design handle. Rather than relying solely on maximising 
the total transition entropy, materials can now be designed to favour a specific microscopic entropy channel, 
vibrational, orientational, or a balanced combination of both, depending on whether pressure-stability or a 
strongly pressure-tunable response is the desired operational target.

Looking forward, the clear and monotonic trends identified in this study suggest that the established 
hydrogen-bond-density/entropy-character relationships may extend to other families of hydrogen-bonded plastic crystals 
beyond the neopentane-derived series studied here, offering a transferable screening criterion for the computational and
experimental search for new barocaloric and thermal-energy-storage materials. Future work combining this entropy-decomposition 
framework with a broader chemical space, including molecules with asymmetric hydrogen-bond donor/acceptor patterns or 
mixed functional groups, could further test the generality of the mechanisms identified here and refine hydrogen-bond
connectivity as a quantitative molecular design principle for tailoring the entropic response of solid-state caloric 
and thermal-energy storage materials.

\section*{Acknowledgments}
The authors acknowledge support by MICIN/AEI/10.13039/501100011033 and ERDF/EU under the grants CNS2025-165467, PID2023-146623NB-I00, 
and the Maria de Maeztu Units of Excellence Programme CEX2023-001300-M, and by the Generalitat de Catalunya under the grant 
2021SGR-00343. J.-Ll. Tamarit and P. Lloveras are grateful to the ICREA Academia program. Computational support was provided by the 
Red Española de Supercomputación under the grants FI-2024-2-0006, FI-2024-1-0012, FI-2024-1-0015 and FI-2025-2-0015. 
\\

\end{document}